\documentclass[sigconf]{acmart}

\AtBeginDocument{%
  \providecommand\BibTeX{{%
    \normalfont B\kern-0.5em{\scshape i\kern-0.25em b}\kern-0.8em\TeX}}}

\setcopyright{acmcopyright}
\copyrightyear{2023}
\acmYear{2023}
\acmDOI{XXXXXXX.XXXXXXX}

\acmConference[CHI '23, Workshop: WS33]{Make sure to enter the correct
  conference title from your rights confirmation email}{April 23--28,
  2023}{Hamburg, Germany}
\acmPrice{15.00}
\acmISBN{978-1-4503-XXXX-X/18/06}

\begin{document}

\newcommand{\ankolika}[1]{{\small\color{blue}{\bf [ankolika: #1]}}}

\title[WOC and Algorithmic Changes]{Instagram versus Women of Color: Why are Women of Color Protesting Instagram’s
Algorithmic Changes?}


\author{Ankolika De}
\email{}
\orcid{}
\affiliation{%
  \institution{College of Information Sciences and Technology, Pennsylvania State University}
  \country{USA}
}
\renewcommand{\shortauthors}{De}

\begin{abstract}
Instagram has been appropriated by communities for several contemporary social struggles, often
translating into real world action. Likewise, women of color (WOC) have used it to protest,
share information and support one another through its various affordances. However, Instagram
is known to have frequent updates, and recently the updates have been more drastic. The newest
update changed the recommendation algorithm such that it showed video-oriented content (reels) from
unknown accounts over static media from a user’s own network. Several marginalized communities, and especially WOC resisted this change
and others that led to it. Due to the backlash, Instagram rolled back its changes. Drawing
from past HCI work on digital platforms for marginalised communities, I propose a qualitative study informed by the open research strategy to understand why WOC are resisting these changes, and eventually provide
implications for design that can help implement changes in a more inclusive manner.
\end{abstract}

\ccsdesc[500]{Human-centered computing~Human computer interaction (HCI)}
\ccsdesc[100]{Human-centered computing~Critical Computing and Social Justice}

\keywords{Social Movements, Online Organizing, Algorithmic Change, Marginalization, Social Media Activism.}

\maketitle

\section{Introduction}
 Several marginalized communities have appropriated Instagram for contemporary social struggles \cite{haq2022} often leading to real world action \cite{Segerberg2011} . In response, the platform needs to evolve to include such communities, as the ramifications of only allowing certain groups to use and benefit from it can be unethical or even dangerous \cite{marjanovic2022theorising, ryu2022microaggression} .

Instagram recently rolled back some changes from its platform, after facing backlash from multiple online communities \cite{huang_isaac_2022} . The most recent update recommended video-oriented content (\emph{Reels}) from unknown accounts to its users, instead of other types of content from their friends network \cite{huang_isaac_2022} . This meant that those who shared static content such as infographics and photos were subdued. Hence, users were unable to see important posts from their own network, unless it was in the \emph{reel} format. WOC were especially resistant to the platform changes and other prior updates, saying that it led to content suppression, wherein some key content was subdued as it wasn’t tailored to be amplified by the recommendation system \cite{iqbal_2020}. As communities use social media for many purposes, ignoring such concerns can increase the current social divide and hamper the digital identities of vulnerable communities \cite{dym2018}. 

Designing online spaces has taken up an important place in Human-Computer Interaction (HCI) research because it goes far beyond simple aesthetics due to its accountability for acceptance, accessibility and inclusivity \cite{stumpf2020, gleason2020} . Frequent design updates on such online platforms are relatively common, as companies attempt to keep up with what the majority wants, and transform into an application for holistic entertainment \cite{clark_2021} . Prior research has investigated users’ reactions on updates in video games \cite{jan2019} and music sharing applications \cite{morreale2020} , finding generally negative feelings as users had trouble adjusting to the new features to perform their old tasks. However, this analysis has never been done for vulnerable communities that use social media platforms for purposes of empowering themselves. Therefore this study aspires to qualitatively answer the following research questions:

RQ1: How do WOC feel about Instagram’s new update, where reels from unknown accounts are shown over content from a friend’s feed?

RQ2: What design changes can be made to Instagram to help WOC use the application better and have more positive experiences?

RQ3: How should future updates be deployed to avoid such resistance and have a smoother transition?


Through careful analysis of the data, I hope to design a protocol for Instagram to introduce updates in a more inclusive manner. The study findings and discussions could be generalized to other platforms that are on a similar move as the insights will have common suggestions to help marginalized users navigate through platform changes.


\section{Related Work}
\subsection{Inequalities on Social Media}
Social media has been compared to a \emph{social system} with \emph{attention} being the currency \cite{Zhu2016AttentionII} . Thus, a difference in the amounts of \emph{attention} gained through social media affordances is said to cause inequalities. 




\subsubsection{Algorithmic Inequalities}
Creators on social media believe that platform governance is implemented unfairly, with some content being more invisible than others \cite{duffy0} . This is consistent with studies that claim social media algorithms are unclear, inconsistent and unstable \cite{bucher2012, savolainen2022} . Unfair decision making \cite{Mewa2020Fairness,feldman_certifying_2015} , biased human implementation \cite{erik2023} , dataset bias \cite{Fazelpour2021-FAZABS} , and intentional ignorance by creators \cite{burrell} are some of the causes reported for these problems. Prior studies have attempted to tackle these, by encouraging the inclusion of public institutions in the process of algorithm design \cite{reviglio2020} , creating affirmative recommendation systems that account for communities that lack a fair representation on datasets \cite{stoica2018} , and reducing the polarizing nature of algorithms to diversify them \cite{sherry2021} . While these are promising ideas, developers are often incentivised with other constraints including revenue generation and global competition, as a result of which battling algorithmic inequalities becomes secondary \cite{siva2020} . 
\subsubsection{Unfairness towards historically marginalized communities}
Researchers have found several cases of bias towards historically marginalized communities due to unfair algorithmic practices \cite{johnson2017, stoica2018}. Haimson et al., (2021) found that content moderation on social media often flagged transgender and black users’ content wrongly, and in the process removed content relevant to expressing their identities, and media articles have supported this claim by reporting the censorship of LGBTQ+ groups on all digital platforms \cite{haimson2021} . Moreover, Are (2020) explored Instagram’s algorithmic bias and proceeded to report that it allowed for online abusers to flourish, while shutting down vulnerable populations \cite{are2020} . Extending to this, a study reported that this results in hate speech, and harassment online, targeted towards women, and other vulnerable populations \cite{golbeck2018}. With the pervasive nature of data-centric practices that focus heavily on recommendations, it has become difficult to stop societal bias translating into data bias, making digital platforms unsafe for marginalized communities \cite{zhao2022}. 


\subsection{WOC on Social Media}
WOC use social media to share opportunities, motivations and highlight problems that are exclusive to their community \cite{brown_2022}. As the online world allegedly allows for more anonymity \cite{lim2011}, it levels the ground for not just activism but for WOC among other vulnerable groups to also experiment with their identities \cite{stamps2022} , form communities \cite{alaoui2015women} and make professional connections \cite{titanji2022} . This section highlights the role WOC play in activism on social media.

\subsubsection{Social Media Activism}
Social media provides affordances such as efficient distribution of information \cite{kou2017one} , dynamic community formation \cite{siddiqui2016social} , quick mobilization and support generation among others \cite{garrett} , that make it an efficient tool for activism. Prior literature has explored the various ways in which this activism was enacted, including but not limited to Black Lives Matter (BLM) \cite{mundt2018} , \#MeToo \cite{manikonda2018}, and The Umbrella movement \cite{lee2015}. WOC were part of several of these intersectional movements, especially playing a major role in the BLM \cite{kia2023} and \#MeToo protests \cite{mueller2021}.  Therefore, social media serves as a tool for empowerment for many vulnerable and marginalized communities including WOC. WOC have also protested against colorism \cite{colorism}, sexism \cite{alex2022}, and other inter-sectional issues using social media. 

\subsubsection{Identity, Community and Support on Social Media}
With the pervasiveness of digital platforms, they are more than just a tool affording social interactions. It now represents a position where users’ \emph{living experiences} are showcased \cite{gunduz2017} and reacted to \cite{gaspar2016}. It helps build support systems \cite{Robinson2016SocialMA}, form communities \cite{waddell2016investigating} and develop identities that are distinct from or similar to their real life versions \cite{sirola2021online}. 


Identity construction is a complex topic in itself and is impacted by self perception \cite{alarid2010}, interactions \cite{page2013stories}, and self motivation \cite{mark1998} . WOC often showcase their identities as a form of rebellion that is different from their stereotypical depictions \cite{Carlson2013TheN} . Similarities perceived amidst constructed identities lead to support systems, which formalize into communities \cite{seargeant2014language} . Users in these communities often lack support systems in their lives, and depend on their online friends to empathize with \cite{ensman2022} . The algorithm plays a major role in this, as it recommends people with similar interests for users to follow \cite{Gillespie2013TheRO} . Changes that modify the recommendation algorithms could hinder the development of communities, as these systems classify similar content to recommend to users and introduce them to relevant communities \cite{stinson} . 


\subsection{Social Media Updates}
 Since their inception, social media applications have been evolving to improve functionality, respond to user feedback, introduce new features and keep up with changes in the global technological infrastructure \cite{Feenberg2009, acker2016} . Some hasty and ill-fitting updates are often a response to market trends and competitive desperation \cite{Feenberg2009} .


Social media changes have never been categorized systematically, but it is relatively easy to understand the two main kinds, after reviewing its evolution. First refer to the structural changes that are explicit and change the way a particular application looks and functions physically. These include user interface (UI) changes that introduce new features or modify old ones \cite{kennedy_2021} . For instance, Twitter rolled out several UI features in 2017 to make the application friendlier for desktop and mobile users \cite {kim} , and Reddit changed the way its search tool was situated to create a more intuitive UI experience for its users \cite{hutchinson_2022} . 

A second category of changes constitute algorithmic modifications, wherein the exact change is not understood, but rather felt in terms of the difference in affordances in the application, which makes it more implicit \cite{devito} . Such is the category of change that Instagram introduced \cite{mclachlan_2022} . As mentioned in the previous sections, algorithms have been blamed for being unfair and biased in terms of their recommendations, content moderation as well as visibility; thus, this change revived feelings of discomfort and unease among users, leading to Instagram overturning them \cite{neuts_2022} .

\subsection{Social Media Design}


Participatory design, and feminist design theory can provide effective design implications for this study. Several scholars have utilised feminist design to propose recommendations for traditional technologies by focusing on interdisciplinary, collaborative and sustainable methodologies \cite{scotford, bardzell2010} . Similarly, participatory techniques have been used to understand the design requirements of groups that have unique experiences \cite{DiSalvo2012CommunitiesPD} . WOC resisting social media changes explain flaws in its implementation that specifically hinder the growth of a particular community and approaching this problem from a more holistic perspective can be helpful and feminist design shows promise here. Thus, the feminist lens with participatory methods can contribute to evaluating a framework for rolling out future changes, and even help design better digital platforms.

\section{Methods}
\subsection{Study Design}
This study aspires to be part of a broader project that focuses on how the constantly changing algorithms of social media impact its users, especially focusing on marginalised communities that use the platform for unique purposes. An open research strategy \cite{bryman2003quantity} was utilized to introduce the present research questions. 

\subsection{Data Collection}
\subsubsection{Interviews}
Users of Instagram who identify as WOC will be recruited via snowball and purposeful sampling \cite{naderifar} , to participate in semi-structured interviews. The interview protocol will be guided by the overarching research questions.  

\subsubsection{Instagram Posts}
For a period of two months, an Instagram search using hashtags such as \#womenofcolor, \#womenofcolorentrepreneurs, \#blackwomen,  and \#brownwomen will be done. Fifty videos per hashtag will be randomly selected to analyze for the study. The  mode of media (video/photo) will be noted. 

\subsection{Data Analysis}
Data collection and analysis will be simultaneous, both for efficiency and to guide data collection as findings emerge. Both modes of data will be analyzed through inductive, iterative coding \cite{bingham2021deductive} , followed by a round of focused coding derived from constructivist grounded theory \cite{charmaz2014constructing} . The first round would proceed with coding the data into specific categories, while the second round would derive more higher level themes. 

\section{Expected Contributions}
As this study contributes to investigating the impacts of frequent algorithmic changes on users, it can help technologists understand how to implement updates in a more inclusive way, such that digital platforms remain efficient in their affordances, especially for marginalized communities. Furthermore, as a stand alone study, it can contribute to understanding how WOC use social media, and the challenges that they face while doing so to provide design implications to build more socially just platforms. Finally, tangential studies can formulate algorithmic designs that are motivated to empower communities by focusing on their needs. Overall, this study can contribute to creating just, fair, accessible and inclusive platforms while also identifying common concerns that WOC have with the current state of social media.



\bibliographystyle{ACM-Reference-Format}
\bibliography{main.bib}


\end{document}